\documentclass[prl,floatfix,twocolumn,superscriptaddress,showpacs,amsfonts,amssymb]{revtex4}
\usepackage{graphicx}
\usepackage{dcolumn}
\usepackage{bm}
\begin{document}
\preprint{APS/123-QED}
\title{Formation of atom wires on vicinal silicon}
\author{C. Gonz\'{a}lez} \affiliation{Facultad de
Ciencias, Departamento de Fisica Teorica de la Materia Condensada,
Universidad Autonoma de Madrid, Madrid 28049, Spain}
\author{P.C. Snijders}
\affiliation{Kavli Institute of NanoScience, Delft University of
Technology, 2628 CJ Delft, The Netherlands}
\author{J. Ortega}
\affiliation{Facultad de Ciencias, Departamento de Fisica Teorica
de la Materia Condensada, Universidad Autonoma de Madrid, Madrid
28049, Spain}
\author{R. P\'{e}rez}
\affiliation{Facultad de Ciencias, Departamento de Fisica Teorica
de la Materia Condensada, Universidad Autonoma de Madrid, Madrid
28049, Spain}
\author{F. Flores}
\affiliation{Facultad de Ciencias, Departamento de Fisica Teorica
de la Materia Condensada, Universidad Autonoma de Madrid, Madrid
28049, Spain}
\author{S. Rogge}
\affiliation{Kavli Institute of NanoScience, Delft University of
Technology, 2628 CJ Delft, The Netherlands}
\author{H.H. Weitering}
\affiliation{Department of Physics
and Astronomy, The University of Tennessee, Knoxville, TN 37996, and
Condensed Matter Sciences
Division, Oak Ridge National Laboratory, Oak Ridge, TN 37831, USA}
\date{\today}

\begin{abstract}
The formation of atomic wires via pseudomorphic
step-edge decoration on vicinal silicon surfaces
has been analyzed for Ga on the Si(112) surface
using Scanning Tunneling Microscopy and Density Functional Theory
calculations.
Based on a chemical potential analysis involving more than thirty
candidate structures and considering various fabrication procedures, it
is concluded that pseudomorphic growth on stepped Si(112),
both under equilibrium and non-equilibrium
conditions,
must favor formation of Ga zig-zag chains rather than linear atom
chains.
The surface is non-metallic and presents
quasi-one dimensional character in the lowest conduction band.
\end{abstract}
\maketitle
Deposition of submonolayer amounts of metal adsorbates on high-index silicon
surfaces often results in the formation or ``self assembly"
of single-domain chain structures or ``atomic wire arrays" \cite{Himpsel01}.
The basic
idea is that metal atoms should preferentially
adsorb at the highly reactive step-edges and form parallel chains
of metal atoms. The transport properties of atomic chains are currently of
great interest because the venerable quasi-particle concept of
Fermi liquid theory is expected to break down
in one dimension.
Photoemission from atomic gold chains on Si(557) \cite{Himpsel02}
and Si(553) \cite{Himpsel03} revealed
dispersive metallic bands that are strongly one-dimensional
but these surfaces turned out to be complicated
reconstructions with large unit cells. The metal atoms
are being incorporated into the terraces rather than at
the step-edges and the
observed metallicity is believed to originate from the substrate's
dangling bonds. Surface reconstructions generally represent a quirk
of nature and it therefore remains difficult to predict
or engineer the formation of useful structures.

A precise definition of interfaces at the atomic scale is usually
best achieved by epitaxial or pseudomorphic ({\it i.e.} strained overlayer)
growth. In this Letter, we investigate the general feasibility of atom wire
formation by pseudomorphic step-edge decoration for the case of
Si(112) with Ga adatoms
\cite{Jung94,Glembocki97,Baski99,Erwin99}.
As with any surface reconstruction, the geometrical configuration of
the atom wires should reflect the balance between
energy gain through passivation of dangling-bond surface states,
and energy cost from the resultant
increase of {\it bond angle strains}. In the case of pseudomorphic
overlayers, one rather speaks of {\it misfit strain},
which can be relieved by forming ``misfit dislocations"  or,
in the case of a monoatomic overlayer, vacancy-line superstructures.

The general question that has not been addressed is whether atomic
chain structures can actually represent
stable or metastable structures
or whether additional metal atoms should passivate
the dangling bonds of the terraces, thereby forming a more complex overlayer.
We present new
Scanning Tunneling Microscopy (STM) data of Si(112)$6\times1$-Ga that reveal
unprecedented detail, including a hitherto unnoticed
symmetry breaking in the vacancy lines, ruling out
the formation of the {\it linear} atom chains that had been proposed before
\cite{Jung94,Glembocki97,Baski99,Erwin99,Yoo02}.
The STM data are complemented with extensive
Density Functional Theory (DFT) calculations and
theoretical STM simulations. Our analysis shows that
Ga atoms
adsorb on the terraces as well as at the step-edges,
forming double chains or
{\it zig-zag} chains.
A chemical potential analysis confirms that the zig-zag
structure represents the thermodynamically stable structure.
The passivation of the substrate's dangling bonds
clearly dictates the structure of the metal overlayer;
the alternative
of creating buckling distortions on the terraces to eliminate
partially-filled dangling bonds
\cite{Yoo02}
appears less effective in lowering the total energy, which precludes
the formation of single-atom wires.
This simple passivation argument should have general validity for
semiconductor interfaces,
which suggests that
single-atom linear wires cannot be stabilized under the
quasi-equilibrium conditions of a
molecular beam epitaxy experiment, and other alternatives ({\it e.g.}
zig-zag chains) must be explored.


Experiments were carried out in an ultra-high vacuum system (base
pressure $<$~5~$\times~10^{-11}$~mbar) with a Ga deposition
source, direct current sample heating facilities and a
variable temperature STM.
The sample was flashed at 1475 K to remove the native
oxide. During resistive heating, the current was directed parallel
to the nano-facets of the clean (112) surface (\emph{i.e.} in the
[1$\overline{1}$0] direction) in order to avoid current-induced
step bunching.
Ga was deposited using a commercial effusion cell.
We have prepared the
Si(112)$6\times1$-Ga
surface in two different ways. In the ``one-step" procedure, Ga is
evaporated onto a Si(112) substrate kept at a temperature of
$T_{s} = 825$~K
\cite{Glembocki97}.
In the ``two-step" procedure, Ga is deposited onto a Si(112) substrate kept
at room temperature. After depositing more than 1 ML, the substrate is
annealed to
$T_{s} = 825$~K to remove excess Ga and form the
$6\times1$ structure \cite{Baski99}.
Experimentally, the one-step and two-step growth procedures always
produced identical $6\times1$ structures.
STM experiments were performed
at room and at low ($\sim$ 40 K) temperature using
etched tungsten tips. STM images of the filled and empty
electronic states were obtained with a constant current between
0.06 and 0.2~nA and bias voltages between 1 and 2~V.


\begin{figure}[ht]
  \centering
\vspace*{0.1cm}
\includegraphics[width=7.5cm]{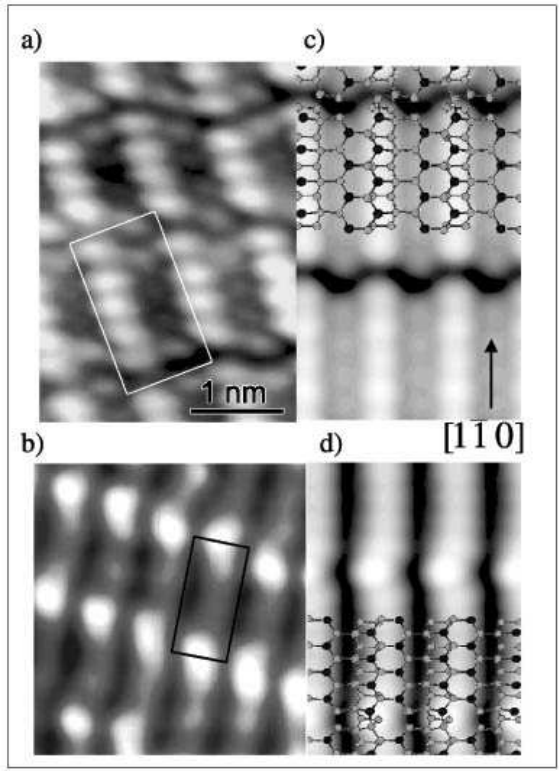}
  \caption{(\emph{a}) Empty state STM image of the Si(112)$6\times1$-Ga
  surface. Tunneling conditions: 1.5 V,  0.2 nA.
  (\emph{b}) Filled state
  STM image. Tunneling
  conditions: -2.0 V, 0.1 nA.
  (\emph{c}) Empty state and
  (\emph{d}) filled state theoretical STM images corresponding to surface
structure of Fig. \ref{fig:strucmod}.}
  \label{fig:stmexp}
\end{figure}

Fig.~\ref{fig:stmexp}(\emph{a}) shows an atomic resolution empty
state STM image of the Si(112)$6\times1$-Ga surface. Two parallel
atomic lines are observed per unit cell, running in the
[$1\overline{1}0$]
direction, intersected by quasi-periodical vacancy lines. The
brightest atomic lines in Fig.~\ref{fig:stmexp}(\emph{a}) are the same
atomic lines
as imaged by Baski \emph{et al.}~\cite{Baski99}. But in addition
we observe an atomic line lying in between the brighter
lines. These two parallel atomic lines appear to form a
\emph{zig-zag} pattern, resulting in a \emph{zig-zag asymmetry} in the
vacancy line.
In the filled state image, Fig.
\ref{fig:stmexp}(\emph{b}), a  relatively bright protusion is the dominant
feature.
Registry aligned dual bias images show that this bright
protrusion
is located in the vacancy lines.
Bright protrusions in different unit cells are connected by fuzzy lines
along the
[$1\overline{1}0$] direction.
These STM measurements show
that the step-edge decorated structural model
\cite{Jung94,Glembocki97,Baski99,Erwin99,Yoo02}
of the Si(112)$6\times1$-Ga surface cannot be correct since the
vacancy lines in that model have mirror plane symmetry ({\it i.e.}
mirror plane perpendicular to
[$1\overline{1}0$]). We furthermore notice that low-temperature STM (40
K) does not provide evidence for a buckling distortion of the Si terrace
atoms, as predicted for the step-edge decorated structure \cite{Yoo02}.
Rutherford
Backscattering Spectrometry experiments indicate
$9 \pm 1$ Ga atoms per $6\times1$ unit cell, as compared to
5 Ga atoms for the step-edge decorated
model.


\begin{figure}[ht]
  \centering
\includegraphics[width=8cm]{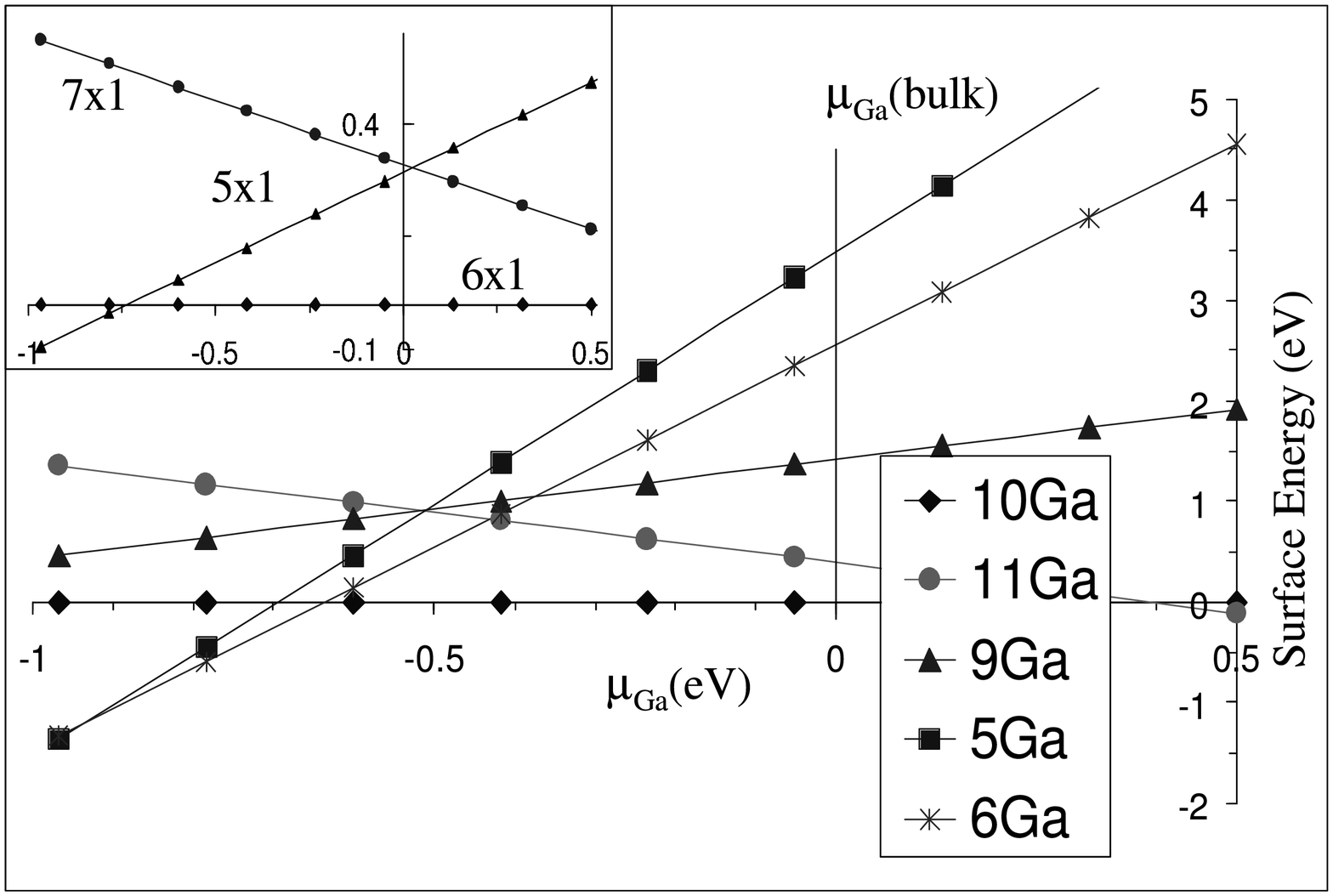}
  \caption{Surface energy as a function of $\mu_{Ga}$ for the most
promising Si(112)$6\times1$-Ga candidate structures.
The structure with 10 Ga atoms per $6\times1$ unit-cell (see Fig.
\ref{fig:strucmod}) is used as reference.
Inset: analysis of the stability of the zig-zag structure (Fig.
\ref{fig:strucmod}) with respect to the longitudinal periodicity.
Note the different vertical scale (also in eV).}
  \label{fig:muga}
\end{figure}

Extensive DFT calculations were performed in order to
identify the precise atomic structure of the Ga/Si(112) surface.
New candidate structural models corresponding to Ga-coverages
ranging from 5 to 11 Ga atoms per $6\times1$ unit cell were explored
using  an efficient local-orbital DFT method \cite{Demkov95}.
In total, more than 30
new structures were fully relaxed,
and their relative stability was analyzed
comparing their surface energies
$F = E_{tot} - \mu_{Si} N_{Si} - \mu_{Ga} N_{Ga}$ :
since different surface structures contain
different numbers of Ga and/or Si atoms we have to consider also the
Ga and Si chemical potentials $\mu_{Ga}$ and $\mu_{Si}$ ($E_{tot}$ is the
total energy per $6\times1$ unit-cell, and $N_{Si}$, $N_{Ga}$ are the
number of Si and Ga atoms per $6\times1$ unit-cell).
For $\mu_{Si}$ we have used the total energy of bulk-Si,
and we have analyzed the stability of the different surface structures
as a function of $\mu_{Ga}$.
The most promising surface structures were further refined using a
plane-wave (PW) DFT code \cite{CASTEP}.
Figure \ref{fig:muga} shows the surface energy
as a function of $\mu_{Ga}$, for the
most stable
surface structures containing 5, 6, 9, 10 and 11 Ga atoms
\cite{structures}.
The chemical potential of bulk-Ga, $\mu_{Ga}(bulk)$,
can be expected to be the upper limit for $\mu_{Ga}$ \cite{Baski99}.
The actual chemical potential $\mu_{Ga}$ of the
overlayer
can be
estimated from
the experimental conditions for
Ga deposition. Consider the one-step process in which
the ($6\times1$)-phase is formed under a Ga flux at a
sample temperature of
$T_{s} = 825$~K.
At this temperature the incoming flux is balanced by a flux of Ga atoms
desorbing from the $6\times1$ surface and the sample can be considered
to be in thermodynamic
equilibrium. Approximating the chemical potential in the effusion cell
by the total energy of bulk-Ga, $\mu_{Ga}(bulk)$,
the overlayer chemical potential can be determined from

\[
\mu_{Ga} = \mu_{Ga}(bulk) - k_B T ln (\frac{p_c}{p_s})
\]
where $p_c$ is the Ga vapor pressure in the effusion cell and $p_s$ the
Ga vapor pressure at the sample. We obtain \cite{pcvsps}
$\mu_{Ga} =
\mu_{Ga-bulk} - 0.32 (0.48)$ eV.


\begin{figure}[ht]
  \centering
\includegraphics[width=8cm]{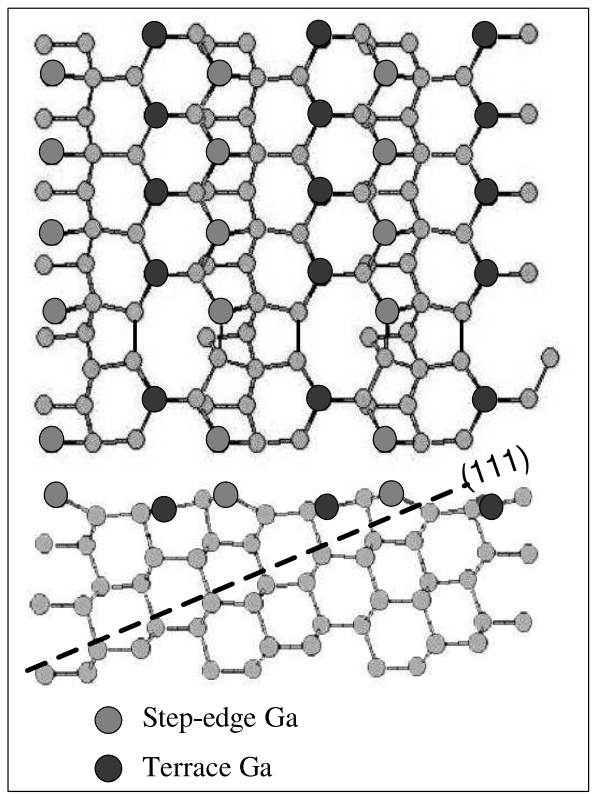}
  \caption{Ball and stick representation of the energy minimized
structure (zig-zag model)
for the Si(112)$6\times1$-Ga surface. Top view and side view.}
  \label{fig:strucmod}
\end{figure}

\begin{figure}[ht]
  \centering
\includegraphics[width=8cm]{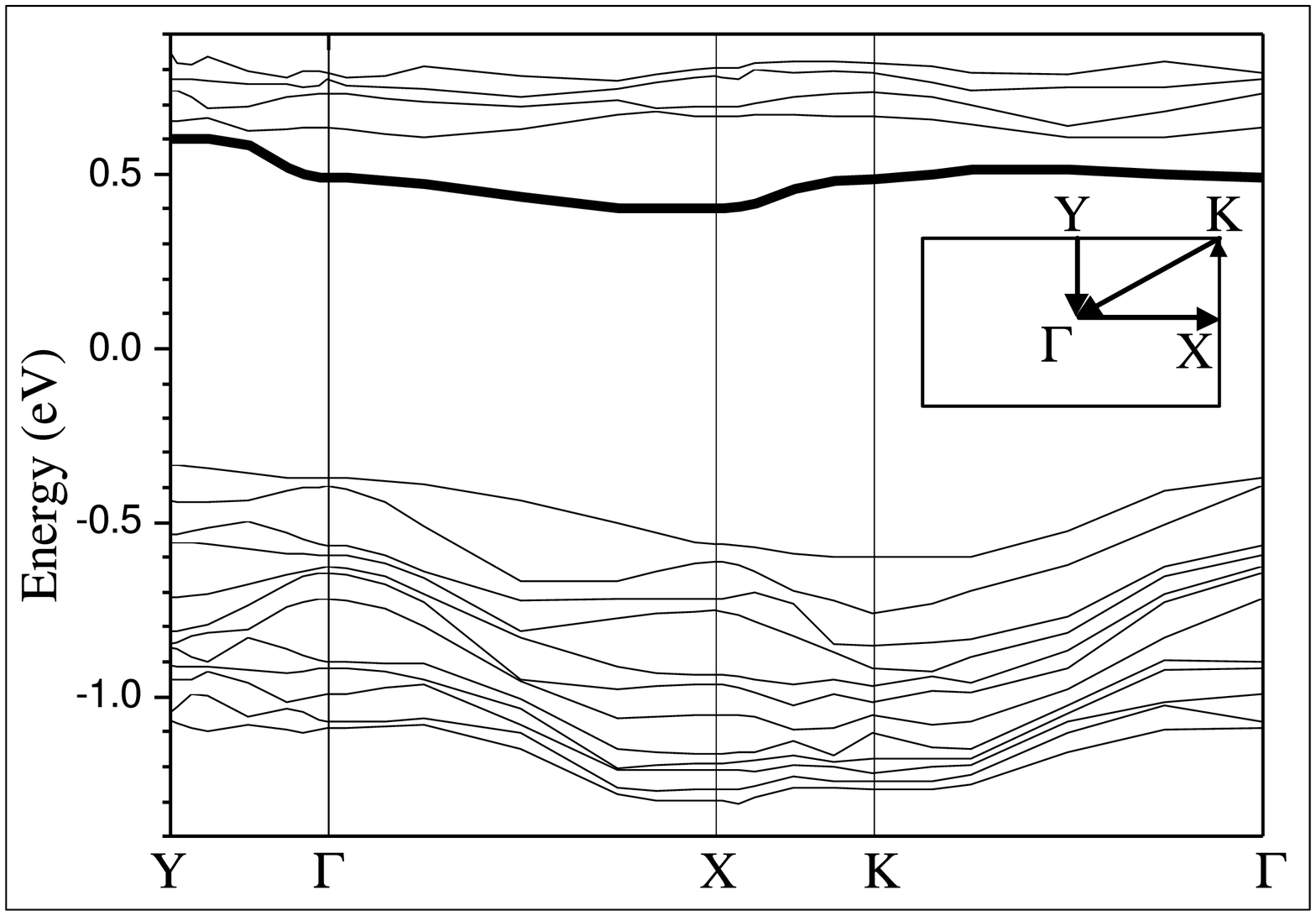}
  \caption{Surface band structure of Si(112)$6\times1$-Ga around the
semiconductor energy gap. The quasi-one dimensional state discussed in
the text is highlighted.  Inset:
two-dimensional Brillouin zone. The $\Gamma$-Y direction corresponds to
the [$1\overline{1}0$] real-space direction.}
  \label{fig:bands}
\end{figure}

In Figure \ref{fig:muga} we observe that the structure with 10 Ga
atoms per $6\times1$ unit-cell is the most stable for this range
of $\mu_{Ga}$. As shown in Fig. \ref{fig:strucmod} the Ga atoms in
this surface form two parallel rows: a row of `step-edge'-Ga atoms
adsorbed at the (111)-like step and a second row of Ga atoms
adsorbed at the terraces. The two Ga rows form a {\it zig-zag}
pattern. Each Ga-row contains 5 Ga atoms per $6\times1$ unit-cell,
{\it i.e.} there is a Ga-vacancy in each row. Inside the vacancy
line a rebonding has taken place, leading to the formation of a
Si-Ga dimer at the step-edge and a Si-Si dimer on the terraces. In
this zig-zag structure all the partially-filled Si dangling bonds
are replaced by empty Ga dangling bonds, and the surface is
semiconducting.
Figure \ref{fig:bands} shows the calculated band-structure for this
surface:
the Ga atoms induce
a surface band
in the upper part of the band gap, with the minimum at the X-point.
The dispersion of this band close to the minimum yields an
effective mass of $m^* \sim 1.48$ along the $X-\Gamma$ direction, while
$m^* \sim 0.15$ along the $X-K$ direction, indicating
quasi-one-dimensional conduction.
This Ga-band is initially empty
but could be populated on heavily-doped n-type subtrates or perhaps by
application
of an external bias.

The step-edge decorated
Si(112)$6\times1$-Ga structure
might be stabilized for very low
$\mu_{Ga}$ values
(5 Ga case in Fig. \ref{fig:muga}).
In order to analyze this possibility we have
calculated the total energy of a clean Si(112) surface \cite{Grein} and
compared the step-edge decorated Si(112)$6\times1$-Ga case with
a hypothetical half-and-half surface:
half of the surface with the zig-zag structure
( {\it i.e.} both step-edge and terrace Ga rows)
and the other half with clean Si(112).
These two cases present the same average Ga coverage.
We find the half-and-half case to be lower in energy by $\sim 0.7$
eV/(10 Ga atoms) for all $\mu_{Ga}$,
showing that the step-edge decorated case is
unlikely to be stabilized at lower coverage.
We also investigated the possibility that the metastable step-edge
decorated structure is stabilized {\it kinetically} through selective
desorption of the Ga terrace atoms during the annealing step of the
two-step process ($\mu_{Ga} \ll \mu_{Ga}(bulk)$). Total energy
calculations, however, show that the desorption energies for
the Ga atoms of Fig.
\ref{fig:strucmod} are lower for step-edge atoms than for
terrace atoms by $\sim$ 0.7 eV/atom. Hence, there seems to be no kinetic
pathway to achieve the step-edge decorated structure via thermal
processing.

The stability of the zig-zag model (Fig. \ref{fig:strucmod})
as a function of the longitudinal periodicity, {\it i.e.}
changing the number of Ga atoms between vacancy lines in each Ga row
from 4 ($5\times1$) to 6 ($7\times1$), is analyzed in the inset of
Fig. \ref{fig:muga}.
For our range of $\mu_{Ga}$ the $6\times1$ periodicity is the most
stable, while the $5\times1$ is only 0.1-0.2 eV/($6\times1$ unit-cell)
higher, in good agreement with the experimental observation of
coexisting
$5\times1$- and $6\times1$-like unit cells on the surface
\cite{Baski99,Yoo02}.
The preferred $6\times1$ periodicity of the vacancy lines appears
identical to that of the simple Frenkel-Kontorova prediction for
one-dimensional chains \cite{Erwin99}.


Figs.~\ref{fig:stmexp}(\emph{c-d})
show simulated STM images \cite{Mingo96}
for the zig-zag model
with a top view
of the structural model superimposed.
The agreement between the experimental and theoretical STM images
is excellent.
It shows that the two atomic lines
imaged in the empty state STM images are indeed the step-edge Ga
line and the terrace Ga line. Furthermore, the zig-zag asymmetry in the
vacancy lines observed experimentally is neatly reproduced by the
simulated STM images.
In the filled state image, fuzzy lines
with a bright protrusion inside the vacancy line are observed, also in agreement
with the experimental images. The bright protrusion originates from the Si-Ga dimer
in the vacancy lines, while the
fuzzy lines are due to Si-Ga bonds on the
(111)-like terrace, close to the step-edge.


\vspace*{0.4cm}

All these results convincingly prove that the
Si(112)$6\times1$-Ga surface contains
a step-edge Ga row {\it and a terrace Ga row}:
the reduction of the number of partially-filled
dangling-bonds is
the main mechanism determining the
atomic structure of the metal overlayer.
In retrospect, the present findings may have come as no surprise.
Chemical passivation should always be the key driving force for
two-dimensional overlayer growth and/or interface reconstruction.
Trivalent Ga is a perfect candidate for passivating a silicon surface
whereas monovalent metals such as Ag and Au generally induce a more
complicated reconstruction \cite{Himpsel01,Himpsel02,Himpsel03,Erwin98}.
This paper, however, highlights a more subtle point, that is, the
thermodynamic and kinetic
stability of these and other atom wire structures should always be addressed
in the context of the applicable range of adsorbate's chemical
potential. Specifically, the fabrication of potentially useful nanowire
structures on stepped surfaces must take into account nature's
desire to passivate dangling bonds on the terraces.
The kinetic limitations could be more case-specific. Our conclusions
should nonetheless have quite general validity and therefore reinforces
the need for similar analysis on other dangling-bond wire arrays,
including those highlighted in Ref. \cite{Himpsel01}, to understand the
generic mechanisms that control the formation of atomic wires on stepped
semiconductor surfaces.


In conclusion, we have analyzed the formation of atom wires on vicinal
semiconductor surfaces for the case of Ga on Si(112).
Detailed STM experiments and
thorough DFT
calculations show that
the passivation of
the substrate's dangling bonds is the main mechanism determining the
structure of the overlayer. The
Si(112)$6\times1$-Ga surface
consists of two parallel rows of Ga
atoms adsorbed on the terraces and at the step-edges, intersected by
quasi-periodic vacancy lines. The surface is non-metallic and
exhibits quasi-one dimensional character
in the lowest conduction band.

This work is sponsored in part by NSF under contract No. DMR-0244570,
the Ministerio de Ciencia y Tecnolog\'{\i}a (Spain) under
grant No. MAT-2001-0665, the Stichting voor Fundamenteel Onderzoek
der Materie and the Royal Netherlands Academy of Arts and Sciences.
We thank T.M. Klapwijk for his stimulating support, and the AMOLF institute
in Amsterdam for the RBS experiments.
Oak Ridge National Laboratory is managed by UT-Battelle, LLC, for
the US Department of Energy under contract No. DE-AC-05-00OR22725.


\begin{thebibliography}{99}
\bibitem{structures}
The 5 Ga case corresponds to the step-edge decorated linear chain of Ga
atoms; the other cases are zig-zag-like structures. Details of these and
other structures will be presented in a forthcoming paper.
\bibitem{pcvsps}
Since the effusion cell flux
is proportional to its vapour pressure, $p_c$, times the cell
aperture area, and the sample flux is also proportional to its
corresponding vapour pressure, $p_s$, times the sample area, we
conclude that $(p_c/p_s) \sim 10^{2} (10^{3})$.
\end{thebibliography}

\end{document}